\documentclass[a4paper]{article}

\usepackage{INTERSPEECH2021}
\usepackage{hyperref}
\usepackage{adjustbox}
\usepackage{xcolor}
\usepackage{algorithm}
\usepackage{algorithmic}
\usepackage{subcaption}
\title{PATE-AAE: Incorporating Adversarial Autoencoder into Private Aggregation of Teacher Ensembles for Spoken Command Classification}
\name{Chao-Han Huck Yang$^1$, Sabato Marco Siniscalchi$^{1,2,3}$, Chin-Hui Lee$^1$}
\address{
  $^1$Georgia Institute of Technology, Atlanta, GA, USA\\
  $^2$Faculty of Computer and Telecommunication Engineering, University of Enna, Italy\\$^3$Department of Electronic Systems, NTNU, Trondheim, Norway}
\email{\{huckiyang,chl\}@gatech.edu, marco.siniscalchi@unikore.it}

\begin{document}

\maketitle
\begin{abstract}
  We propose using an adversarial autoencoder (AAE) to replace generative adversarial network (GAN) in private aggregation of teacher ensembles (PATE), a solution for ensuring differential privacy in speech applications. The AAE architecture allows us to obtain good synthetic speech leveraging upon a discriminative training of latent vectors. Such synthetic speech is used to build a privacy-preserving classifier when non-sensitive data is  not sufficiently available in the public domain. This classifier follows the PATE scheme that uses an ensemble of noisy outputs to label the synthetic samples and guarantee $\varepsilon$-differential privacy (DP) on its derived classifiers.
  Our proposed framework thus consists of an AAE-based generator and a PATE-based classifier (PATE-AAE). Evaluated on the Google Speech Commands Dataset Version II, the proposed PATE-AAE improves the average classification accuracy by +$2.11\%$ and +$6.60\%$, respectively, when compared with alternative privacy-preserving solutions, namely PATE-GAN and DP-GAN, while maintaining a strong level of privacy target at $\varepsilon$=0.01 with a fixed $\delta$=10$^{-5}$.
\end{abstract}

\section{Introduction}
The speech signal contains a rich set of information~\cite{pathak2013privacy} that encompasses gender, accent, speaking environment, and other speaker characteristics; therefore, protecting data privacy becomes a raising concern when speech data is used to deploy commercial speech applications. In recent years, public regulations, e.g., GDPR~\cite{voigt2017eu} and CCPA~\cite{shatz2020california}, have been proposed to establish new guidelines related to data privacy measurement and identity protection in end-user applications. Recent works on model inversion attacks~\cite{fredrikson2015model, carlini2020extracting} indeed highlighted the importance of data privacy when the original data profile (e.g., facial images~\cite{fredrikson2015model}) could be recovered from a machine learning model by using query-free optimization techniques.

Differential privacy~\cite{dwork2008differential} (DP) is an effective mechanism for ensuring individual data protection, and it has been deployed in several industrial systems~\cite{abadi2016deep, kifer2020guidelines}\footnote{Apple has also applied differential privacy with a privacy budget ($\varepsilon$=8) based on an official document in \url{https://www.apple.com/privacy/docs/Differential_Privacy_Overview.pdf}.} to protect customer's sensitive information by exploiting a sophisticated noisy perturbation scheme. The $\varepsilon$-DP mechanism~\cite{dwork2008differential} provides a way to quantify a privacy loss and set up a privacy budget (e.g., a minimum $\varepsilon$ value) for a given dataset. However, $\varepsilon$-differential private models~\cite{abadi2016deep} need to be refined in order to improve a degraded prediction accuracy~\cite{rajkumar2012differentially} caused by the DP noise. The private aggregation of teacher ensembles~\cite{papernot2016semi} (PATE) is a recently proposed solution that aims to combat the accuracy loss of the machine learning models while ensuring privacy requirements. PATE follows a teacher-student architecture~\cite{ hu2020relational}, where the teacher is an ensemble model. The underpinning idea in PATE is to leverage upon noisy outputs of aggregated teacher models to (re)label non-sensitive public data with DP guarantees. The PATE method and its improved version~\cite{papernot2018scalable} were proven useful in reducing the model accuracy drop through a voting process during the noisy ensemble. Nonetheless, the teacher-student learning process highly depends on a hypothesis~\cite{papernot2016semi, papernot2018scalable, jordon2019pate} that there exists a sufficient amount of public (non-sensitive) data to train the model. PATE-GAN~\cite{jordon2019pate} tries to overcome this issue by incorporating a generative block jointly trained with the PATE block; the goal is providing enough synthetic data to train deep models effectively. Unfortunately, PATE-GAN does not work well for high dimensional data synthesis (e.g., images), as demonstrated in recent studies~\cite{chen2020gs, haque2020high}. Moreover, generating speech samples is a challenging task, as shown in recent studies about neural vocoders \cite{rethage2018wavenet, oord2016wavenet}.%

\begin{figure}[ht!]
    \centering
    \includegraphics[width=0.90\linewidth]{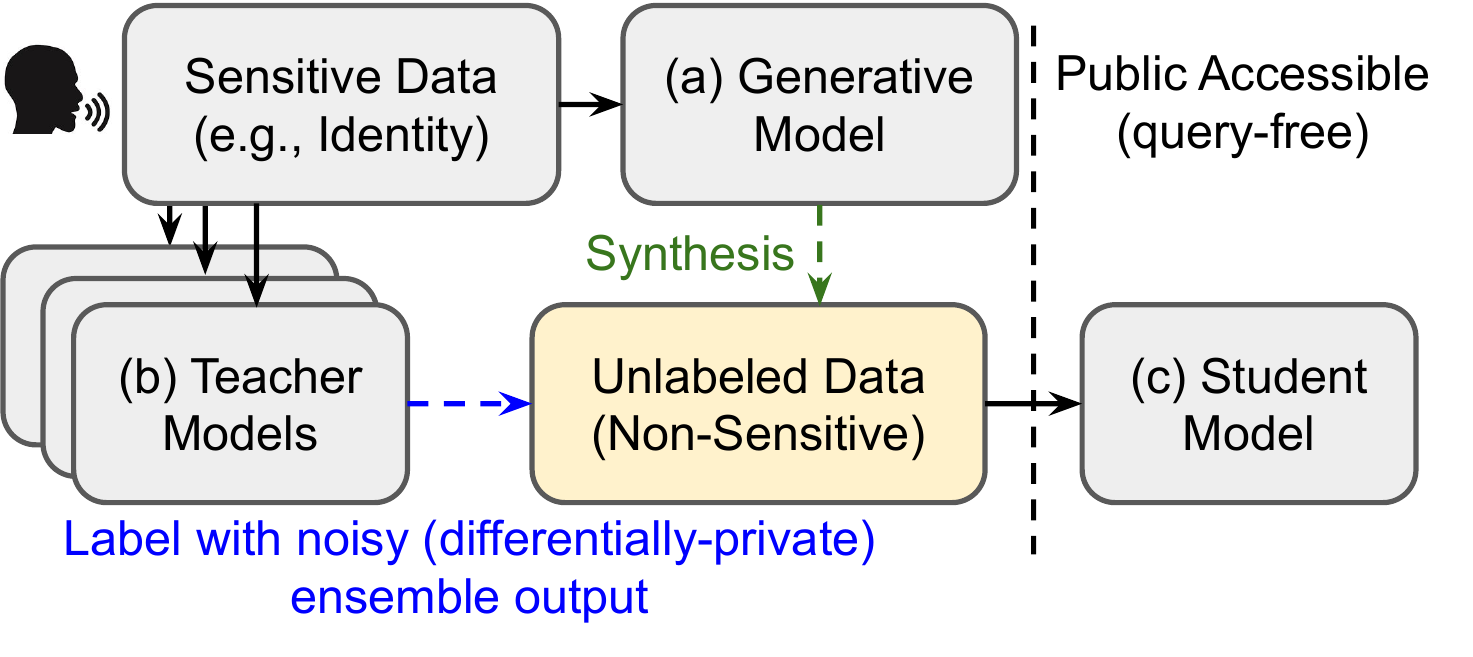}
    \caption{Private aggregation of teachers ensemble (PATE) learning process~\cite{papernot2016semi, papernot2018scalable}: (a) the teacher prediction models training from sensitive data; (b) a joint generative model (e.g., adversarial autoencoder~\cite{makhzani2015adversarial} for audio synthesis in this study); (c) student prediction model training from non-sensitive data. }
    \label{fig:1}
\end{figure}

In this study, we introduce an adversarial autoencoder~\cite{makhzani2015adversarial} (AAE) based model into PATE to improve the generative process for privacy-preserving speech classification. PATE-AAE first adapts an autoencoder to minimize a reconstruction loss, training on sensitive data. As shown in Fig.~\ref{fig:1}(a), the generative model produces synthetic data as non-sensitive samples. Meanwhile, the training data are divided into $I$ isolated subsets to train individual teacher classifiers. For instance, $I$ is equal to $3$ in Fig.~\ref{fig:1}(b). The teacher classifiers then undergo an output aggregation process to generate noisy labels, which ensures the $\varepsilon$-differentially private protection. Finally, a student classifier uses the labeled synthetic samples (non-sensitive data) for training its model. The proposed PATE-AAE framework is assessed with the Google Speech Commands Dataset Version II \cite{warden2018speech}. Our experimental evidence demonstrates competitive results in terms of synthetic sample quality and classification accuracy with a strong $\varepsilon$-DP guarantee ($\varepsilon<1$) considering established privacy-preserving learning (PPL) works~\cite{jordon2019pate, xie2018differentially}. To the best of the authors' knowledge, this is the \textbf{first} attempt to introduce the PATE architecture into a speech classification task. Moreover, the proposed solution is benefited from adversarial autoencoder block, with advantages over existing GAN solutions \cite{fredrikson2015model, jordon2019pate} of having a better test-likelihood estimation.

\section{Related Work}
Much of research effort to preserve data privacy in a machine learning model can be categorized into one of the two main groups: (i) systemic, such as federated learning \cite{leroy2019federated}, data isolation \cite{yang2020decentralizing}, and data encryption \cite{glackin2017privacy}, and (ii) algorithmic, mainly differential private machine learning \cite{abadi2016deep}. In the following sections, we first briefly discuss some of the privacy-preserving solutions proposed for speech applications. Next, we describe the substratum of differential privacy devised for machine learning applications and discuss the difference with our proposed approach while highlighting its key contributions.
\subsection{Privacy-Preserving Speech Processing}
Federated architectures~\cite{leroy2019federated,  dimitriadis2020federated, yang2020decentralizing} have been studied in the speech processing community to increase privacy protection. For example, the average gradient method~\cite{dimitriadis2020federated} was used to update the learning model for decentralized training~\cite{qi2020submodular}. Heterogeneous computing architectures~\cite{yang2020decentralizing, chen2021federated} shows advantages on acoustic feature extraction for vertical federate learning. However, those approaches at a system-level usually make some assumptions on the limited accessibility of the malicious attackers and provide less universal measures about the privacy guarantees. There exist also some algorithmic efforts on investigating privacy-preserving speech processing by using cryptographic encryption~\cite{glackin2017privacy, brasser2018voiceguard}, and computation protocols~\cite{pathak2012privacy}. Meanwhile, these encryption algorithms and protocols barely cover the training sample-level privacy protection, which plays a major role in deploying large-scale machine learning models.

However, differentially private algorithms~\cite{abadi2016deep, dwork2010boosting} is with a different focus from the aforementioned frameworks, aiming to provide quantitative guarantees and further prevent identity (e.g., accent) inference. We define a mathematical formation of differential privacy and investigate potential impacts on speech processing in the following sections.

\subsection{Differential Privacy Fundamentals}
The differential privacy mechanism~\cite{dwork2008differential} is an established standard to deploy algorithms with a target privacy guarantee.

\textbf{Definition 1.} A randomized algorithm $\mathcal{M}$ with domain $\mathcal{D}$ and range $\mathcal{R}$ is $(\varepsilon, \delta)$-differentially private if for any two neighboring inputs (e.g., acoustic data) $d, d^{\prime} \in \mathcal{D}$ and for any subset of outputs (e.g., labels) $S \subseteq \mathcal{R}$, the following holds:
\begin{equation}
\operatorname{Pr}[\mathcal{M}(d) \in S] \leq e^{\varepsilon} \operatorname{Pr}\left[\mathcal{M}\left(d^{\prime}\right) \in S\right]+\delta.
\label{eq:1}
\end{equation}
The above definition provides a notion of privacy that can be interpreted as a measure of the probabilistic difference of a specific outcome by a multiplicative factor, $e^{\varepsilon}$, and an additive amount, $\delta$. Both $\varepsilon$ and
$\delta$ should be positive or equal to zero. Considering $\delta\rightarrow$ 0 with only minor relaxation, a smaller value of $\varepsilon$ indicates a stronger $(\varepsilon, 0)$-differentially private guarantee. In other words, nearly equal probabilities in Eq.~\ref{eq:1} would be given from the neighboring inputs $d$ and $d^{\prime}$, which makes data identity much hard to be inference. Moreover, learning from post-processing features (e.g., mel-frequency cepstral coefficients (MFCC)) from the data could also be differentially private, which have been proofed by the theorem given in~\cite{dwork2008differential}.

\subsection{Differential Privacy for Machine Learning}

More recently,  Abadi \emph{et al.} \cite{abadi2016deep} introduced the composition theorem~\cite{dwork2010boosting}, which guarantees the validity of DP protection when batch-wise training is used to learn deep neural network (DNN) parameters with non-convex objective functions. DP-GAN~\cite{xie2018differentially} incorporated noisy perturbations into a generative model during gradient updates to satisfy $\varepsilon$-DP but shows degraded prediction accuracy. Recent advances in teacher-student ensembles methods, such as PATE~\cite{papernot2016semi, papernot2018scalable}, have further shown state-of-the-art performance on large-scale image classification tasks with sufficient non-sensitive data. Jordon \emph{et al.}~\cite{jordon2019pate} instigated a GAN-based~\cite{goodfellow2014generative} generator into PATE to extend using scenarios with sufficient synthetic data (as non-sensitive), which is called PATE-GAN~\cite{jordon2019pate}. PATE-GAN is aligned with our motivation, but it only shows a stable performance for a small amount and low dimensional data.
Meanwhile, application of PATE to large-scale speech processing, such as spoken-term classification,
is practically absent despite the sensitive nature of the speech signals.
In this study, we propose an autoencoder based approach incorporating PATE into speech processing, which considers non-sensitive acoustic data is not accessible. We will introduce an adversarial autoencoder with PATE to ensure $\varepsilon$-DP for the acoustic modeling in the next section.

\section{PATE-AAE Framework}
The proposed method consists of AAE and PATE models with feature encoders~\cite{chorowski2019unsupervised}. We focus on the application of PATE for speech processing, which is often in shortage of non-sensitive human voice data and more severe than in the original PATE~\cite{papernot2016semi}. It should be noted that PATE-GAN has succeeded in synthesizing low dimensional data (e.g., short sequences of EEG) but has failed when dealing with high-dimension data (e.g., images). This could be due to its difficulties of using a random noise generator to match \emph{input data} distribution from \emph{sample} discriminator in standard GAN~\cite{goodfellow2014generative} training.

\subsection{Private Aggregation of Teacher Ensembles}

We describe the foundation~\cite{papernot2016semi, papernot2018scalable} of PATE to empower privacy-preserving speech classification. First, an ensemble of teacher models is built by partitioning the training dataset of images into $n$ disjoint subsets: $\mathcal{D}_1 ,..., \mathcal{D}_I$. Next, each subset is used to train $I$ classifier independently: $\mathcal{T}_1,..., \mathcal{T}_I$, that is, the teacher models. For each input data $x$, we aggregate prediction outputs from the teacher models to a single prediction.
The number, $c_{j}(x)$, of teachers, that output class $j$ for the given input $x$ with $m$ possible classes is set to be:
\begin{equation}
c_{j}(x)=\left|\left\{\mathcal{T}_{i}: \mathcal{T}_{i}(x)=j\right\}\right| \text { for } j=1, \ldots, m.
\label{eq:2}
\end{equation}
A random perturbation was introduced into the vote count, $c_j$, in Eq.~(\ref{eq:2}) to obtain a noisy final prediction:
\begin{equation}
F_{\operatorname{PATE}}(\mathbf{x},{\lambda})=\underset{j \in[m]}{\arg \max }\left(c_{j}(\mathbf{x})+Y_{j}({\lambda})\right),
\label{eq:3}
\end{equation}
where $Y_{1}, ..., Y_{m}$ are i.i.d. $Lap(\lambda)$ random variables with location $0$ and scale $\lambda^{-1}$. $\lambda$ refers to a privacy parameter that influences $(\epsilon, \delta)$-differentially private guarantees and has been proven its bounded properties under composition theorems applying for model aggregation in ~\cite{papernot2016semi, papernot2018scalable}. As shown in Figure~\ref{fig:1}, the next step in the PATE mechanism is based on a knowledge transfer process, where the noisy ensemble output is used to relabel a non-sensitive dataset, having a total sample number equal to $K$, which in turn is used to train a student model, $\mathcal{S}$. Both prediction outputs and the trained student model's internal parameters are free from querying requests, which allows the privacy cost only associated with acquiring the training data for the student model. Under the aforementioned data setup, the  student model is $(\varepsilon, 10^{-5})$-differentially private guarantee using $\lambda=\frac{K}{2\varepsilon}$ from the analysis in ~\cite{papernot2016semi, papernot2018scalable}. According to Eq.~(\ref{eq:3}), a large $\lambda$ refers to a smaller $\varepsilon$ providing a strong privacy guarantee but degrade the accuracy of the labels from the noisy maximum prediction output of the PATE function.

\begin{figure}[ht!]
    \centering
    \includegraphics[width=0.95\linewidth]{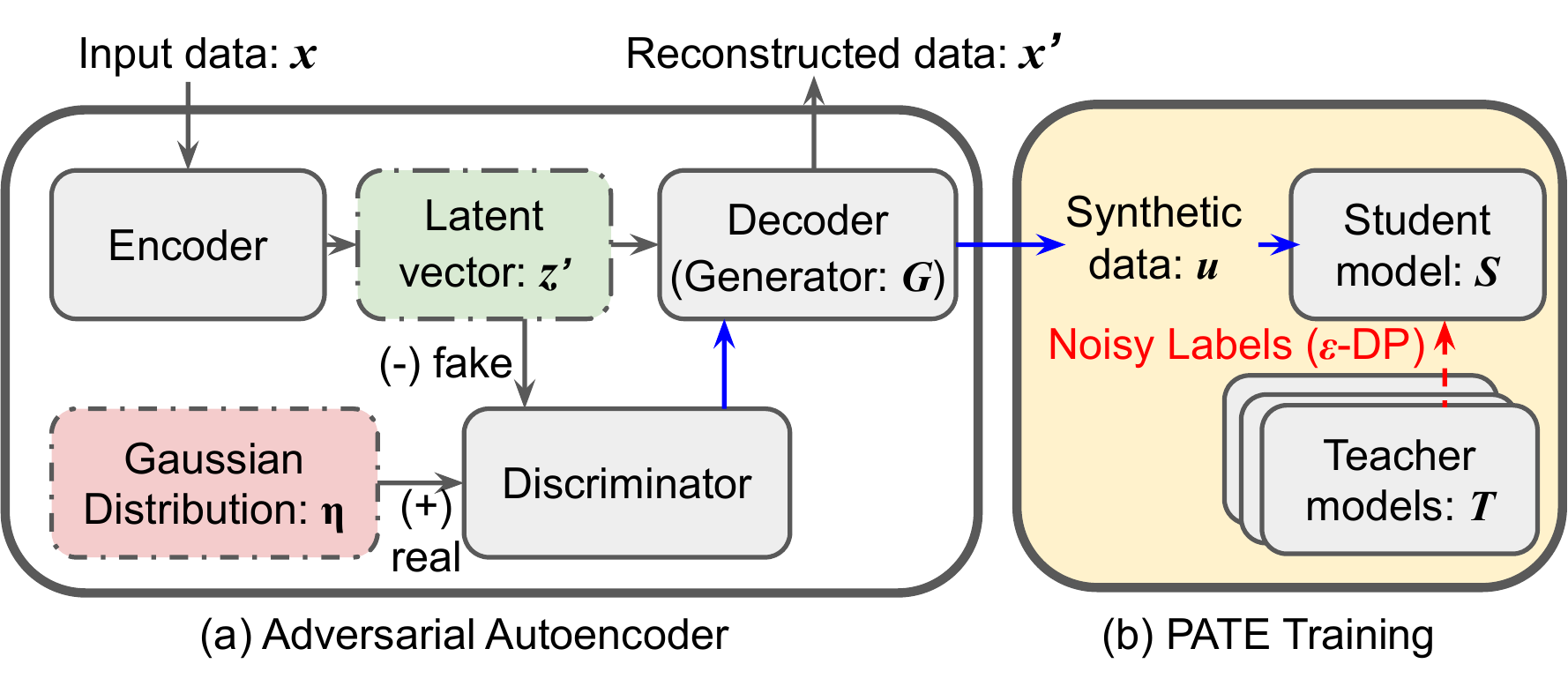}
    \caption{The proposed PATE-AAE framework. }
    \label{fig:2}
\end{figure}

\subsection{PATE with Adversarial Autoencoder (PATE-AAE)}
We introduce AAE training as follows. Instead of training the noisy generator as in GAN, AAE leverages upon an autoencoder-based regression model to minimize a reconstruction loss between input data ($x$) and decoded output ($x^{\prime}$). The bottleneck vector (latent space $z$) is modeling as variational autoencoder~\cite{kingma2013auto} but uses a discriminator to refine $z$ closing to a real vector ($\eta$) sampling from a fixed Gaussian distribution as shown in Fig.~\ref{fig:2}(a). This discriminative training resulted in a better test-likelihood on synthetic samples~\cite{makhzani2015adversarial}. The recent success~\cite{chorowski2019unsupervised} of probabilistic autoencoder for audio synthesis motivates our solution that combines PATE with AAE for spoken command classification. Let $x$ and $z$ be the input and the bottleneck latent vector of an encoder-decoder model, respectively. The universal approximator posterior $q(z)$ introduced \cite{makhzani2015adversarial} is:
\begin{align}
q(z \mid x)=\int_{\eta} q(z \mid x, \eta) p_{\eta}(\eta) d \eta \\
\Rightarrow q(z)=\int_{x} \int_{\eta} q(z \mid x, \eta) p_{d}(x) p_{\eta}(\eta) d \eta d x,
\end{align}
where the stochasticity in $q(z)$ comes from both the data-distribution $x$ and the random noise $\eta$ with a fixed Gaussian distribution at the input of the encoder. The adversarial training procedure can match $q(z)$ to $p(z)$ by back-propagation through the encoder network directly. The encoder encodes an input data $x$ into latent vector, $z^{\prime}_i \sim \mathcal{N}\left(\mu_{i}(\mathbf{x}), \sigma_{i}(\mathbf{x})\right)$, by variational inference used in~\cite{kingma2013auto}. Therefore, a training objective for reconstructing input $x$ is computed by minimizing the following upper-bound on the negative log-likelihood of $x$:
\begin{equation}
E_{x}\left[\mathrm{E}_{q(z \mid \mathbf{x})}[-\log (p(x \mid z)]]+E_{x}[\operatorname{KL}(q(z \mid x) \| p(z))].\right.
\label{eq:6}
\end{equation}

Makhzani \emph{et al.}~\cite{makhzani2015adversarial} further introduce a discriminative update into the second terms of Eq.~(\ref{eq:6}) that makes $q(z)$ to match to the distribution of $p(z)$ to train an AAE. The discriminative training objective between the latent vector $z^{\prime}$ (denoted as a fake sample as $0$) and the sampling noise $\eta$ (denoted as a real sample as $1$) is computed by BCE loss and back-propagated gradients to input data ($x$) for updating encoder's parameters (Fig.~\ref{fig:2} (a)).

Next, for training teacher models, we partition the sensitive dataset into $n$ subsets, $\mathcal{D}_{1}, \ldots, \mathcal{D}_{I},$ with $\left|\mathcal{D}_{i}\right|=\frac{|\mathcal{D}|}{I}$ for $\forall q$. Each teacher model ($T_i$) is training with discriminator loss:
\begin{equation}
\mathcal{L}_{T_i}=-(\sum_{} \log T_{i}\left(q(z) \right)+\sum_{j=1}^{I} \log \left(1-T_{i}\left(q\left(z{'}_{j}\right) \right)\right))
\label{eq:7:tea}
\end{equation}
To generate synthetic samples for training student model, we take $I$ samples of Gaussian distribution, $\eta_{1}, ..., \eta_{I}$ using the trained AAE decoder network ($G$) to synthesize sample $\hat{\mathbf{u}}_{j}=G\left(\eta_{j}\right)$
for each class. Following the PATE mechanism for knowledge transfer, the aggregated noisy output from the teachers models in Eq.~(\ref{eq:7:tea}) labels the synthetic data for training a differentially private student model, where noisy label refers $r_{j}=\operatorname{PATE}\left(\hat{\mathbf{u}}_{j}, \lambda\right)$ from Eq.~(\ref{eq:3}). Finally, we train the student model to maximize the standard cross-entropy loss on this teacher-labeled data:
\begin{equation}
\mathcal{L}_{S}=\sum_{j=1}^{I} r_{j} \log S\left(\hat{\mathbf{u}}_{j} \right)+\left(1-r_{j}\right) \log \left(1-S\left(\hat{\mathbf{u}}_{j} \right)\right)
\end{equation}
We train $G$, $T_{1},...,T_I$ and $S$ iteratively, with each iteration of G consisting of first performing gradient updates on all teachers, then performing gradient updates of the student. The major difference between proposed PATE-AAE and PATE-GAN~\cite{jordon2019pate} is on the generative process. Proposed AAE method uses a regression autoencoder architecture to reconstruct the input samples and use a random variable for the refined latent space as the decoder (generator) input for generating new synthetic samples. Instead, PATE-GAN adapts discriminator on the sample generator directly to refine the learning process from random noise to synthetic data, which produces worse test-likelihood on high dimensional data from previous studies~\cite{chen2020gs, chen2020gan} related to the convergence properties~\cite{kodali2017convergence} of GAN. To conduct privacy-preserving speech processing frameworks, we select PATE-GAN~\cite{jordon2019pate}, and DP-GAN~\cite{xie2018differentially} as baselines in our studies motivated by the condition of without any available non-sensitive audio dataset.

\section{Experiment}
\subsection{Experimental Setup}
\begin{figure}[ht!]
    \centering
    \includegraphics[width=0.99\linewidth]{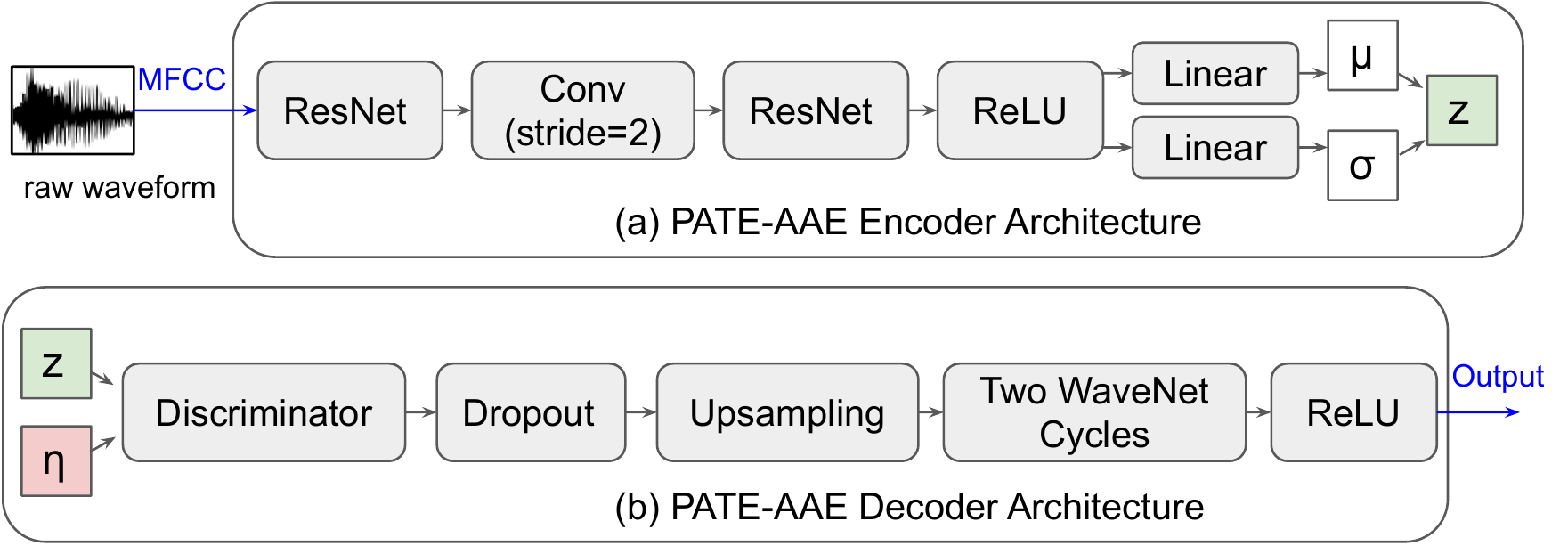}
    \caption{PATE-AAE encoder-decoder architecture. }
    \label{fig:a}
\end{figure}

\textbf{(1) Dataset and Classifier Model:} A large-scale dataset ($\sim$100k training samples)  for the partition process is needed in our experiments. Therefore, we have chosen the Google Speech Commands V2~\cite{warden2018speech} task, which contains 105,829 utterances of 35 words from 2,618 speakers with a sampling rate of 16 kHz. The audio length per sample clip is 1 second for a total amount of $55.5$ hours. We split the dataset into $I$=200~\cite{jordon2019pate} disjoint subsets to train individual teacher classifiers following the procedure indicated in Eq.~(\ref{eq:7:tea}). We use the mel-spectrogram feature with an 80-band mel-scale and 1024 points of discrete Fourier transform as inputs to the classifiers. For a fair comparison, both teacher and student classifiers use an identical self-attention~\cite{vaswani2017attention} and U-Net~\cite{ronneberger2015u} based neural network proposed in the recent work~\cite{yang2020decentralizing}, which has shown benchmark prediction accuracy on the selected speech commands dataset.
\\
\textbf{(2) Encoder-Decoder Model:}
For encoder-decoder inputs, we use standard 13 MFCC at a sampling rate of 100 Hz from 80 log-mel filter-bank features on training ($x$) and synthetic ($u$) data. We carefully build our encoder-decoder upon a WaveNet-based autoencoder presented in~\cite{chorowski2019unsupervised}, which has shown competitive performance for unsupervised speech synthesis tasks. As shown in Fig~\ref{fig:a}(a), our encoder has four network blocks associated 768 input units, which include the first ResNet layer, a convolution layer with a stride scale = 2, the second ResNet layer, and a Dense layer with ReLU activation. The latent representation $z$ is computed by outputs ($\mu, \sigma$) from two linear layers with 128 units. As shown in Fig~\ref{fig:a}(b), our decoder applied a randomized dropout layer from an output of the discriminator, which is selected from a latent vector ($z$) or a random vector ($\eta$). The output vector is then upsampled 320 times to fit the 16 kHz sampling rate for WaveNet decoding. Finally, we follow the same setting with~\cite{chorowski2019unsupervised} for running two cycles of WaveNet with 20 convolution layers followed by a 256-ReLU layer. We follow $\mu$-law companding transformation~\cite{recommendation1988pulse, oord2016wavenet} with 256 quantization levels to generate raw 16k Hz audio.
\\
\textbf{(3) Synthesis and Classification Evaluation Metrics:}
For the speech synthesis task, the Frechet Inception Distance (FID)~\cite{shmelkov2018good} is selected to evaluate the sample quality that computes the Frechet Distance~\cite{dowson1982frechet} between two multivariate Gaussian distributions for the synthetic and real samples. We follow a standard FID setup in~\cite{haque2020high} to evaluate the quality of over 10,000 synthetic speech samples generated from random noise. For the speech command classification task, classification accuracy is used to evaluate the student model. We train each privacy-preserving model 20 times and report its average prediction accuracy.

\begin{table}[ht!]
\centering
\caption{FID scores (lower is better) for speech synthesis quality by generative models with $\varepsilon$-DP (a fixed $\delta$=10$^{-5}$) settings.}
\label{tab:1:fid}
\begin{adjustbox}{width=0.47\textwidth}
\begin{tabular}{|l|ccccc|}
\hline
privacy target ($\varepsilon$) & 0.01 & 0.1  & 1    & 10 & 100   \\ \hline \hline
DP-GAN~\cite{xie2018differentially}         & 35.2$\pm$0.5 & 32.2$\pm$0.6 & 28.8$\pm$0.4 & 26.7$\pm$0.3 & 24.9$\pm$0.3 \\ \hline
PATE-GAN~\cite{jordon2019pate}       & 33.4$\pm$0.5 & 30.1$\pm$0.4 & 27.9$\pm$0.3 & 26.0$\pm$0.3  & 24.3$\pm$0.1 \\ \hline
PATE-AAE       & \textbf{30.2}$\pm$0.4 & \textbf{28.6}$\pm$0.3 & \textbf{26.3}$\pm$0.2 & \textbf{24.7}$\pm$0.2& \textbf{24.0}$\pm$0.1 \\ \hline
\end{tabular}
\end{adjustbox}
\end{table}

\begin{figure}[ht!]
        \centering
        \begin{subfigure}[b]{0.213\textwidth}
            \centering
            \includegraphics[width=\textwidth]{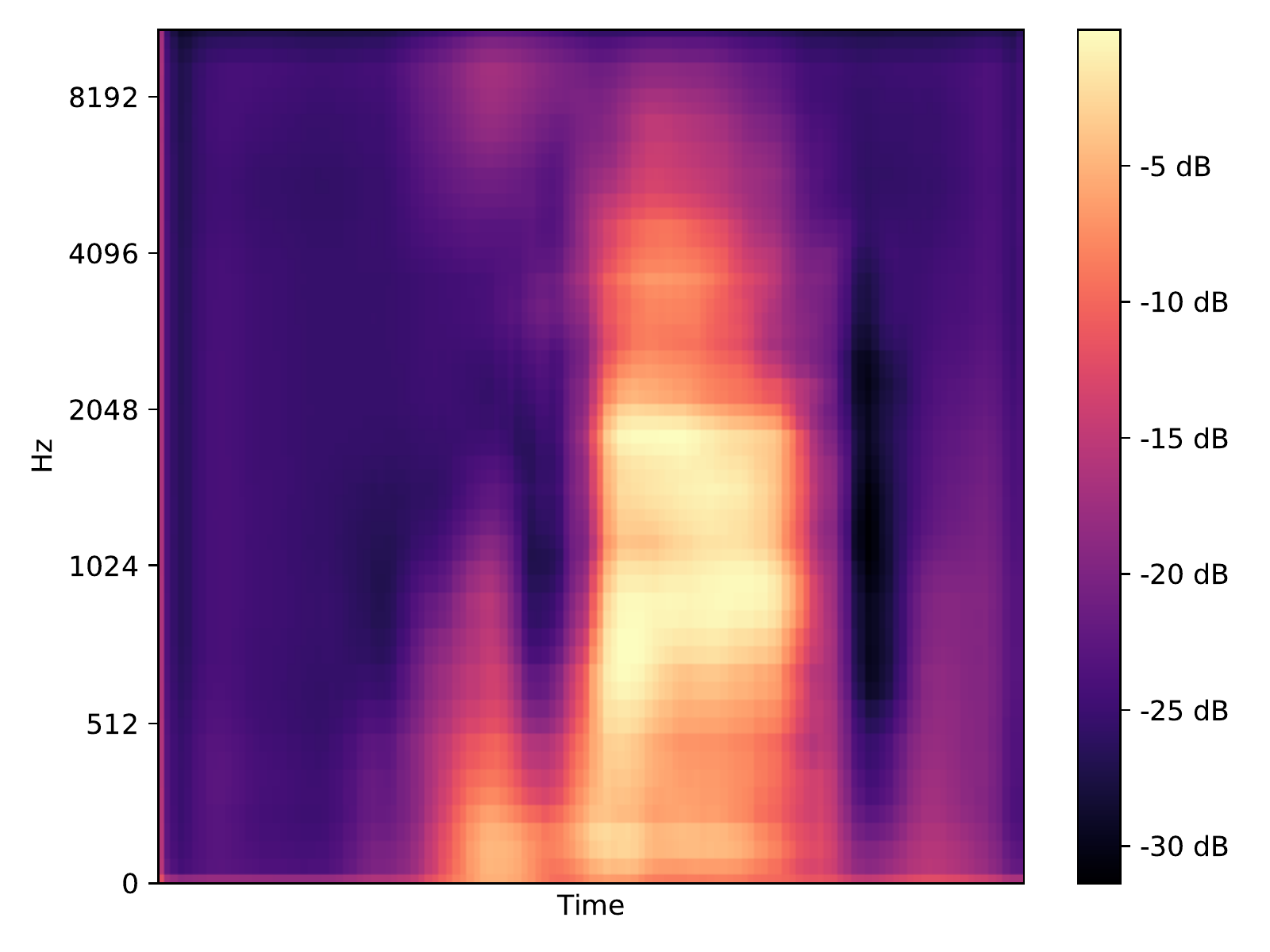}
            \caption[]%
            {{\small PATE-GAN}}
            \label{fig:pateg:s}
        \end{subfigure}
        \quad~~~~
        \begin{subfigure}[b]{0.213\textwidth}
            \centering
            \includegraphics[width=\textwidth]{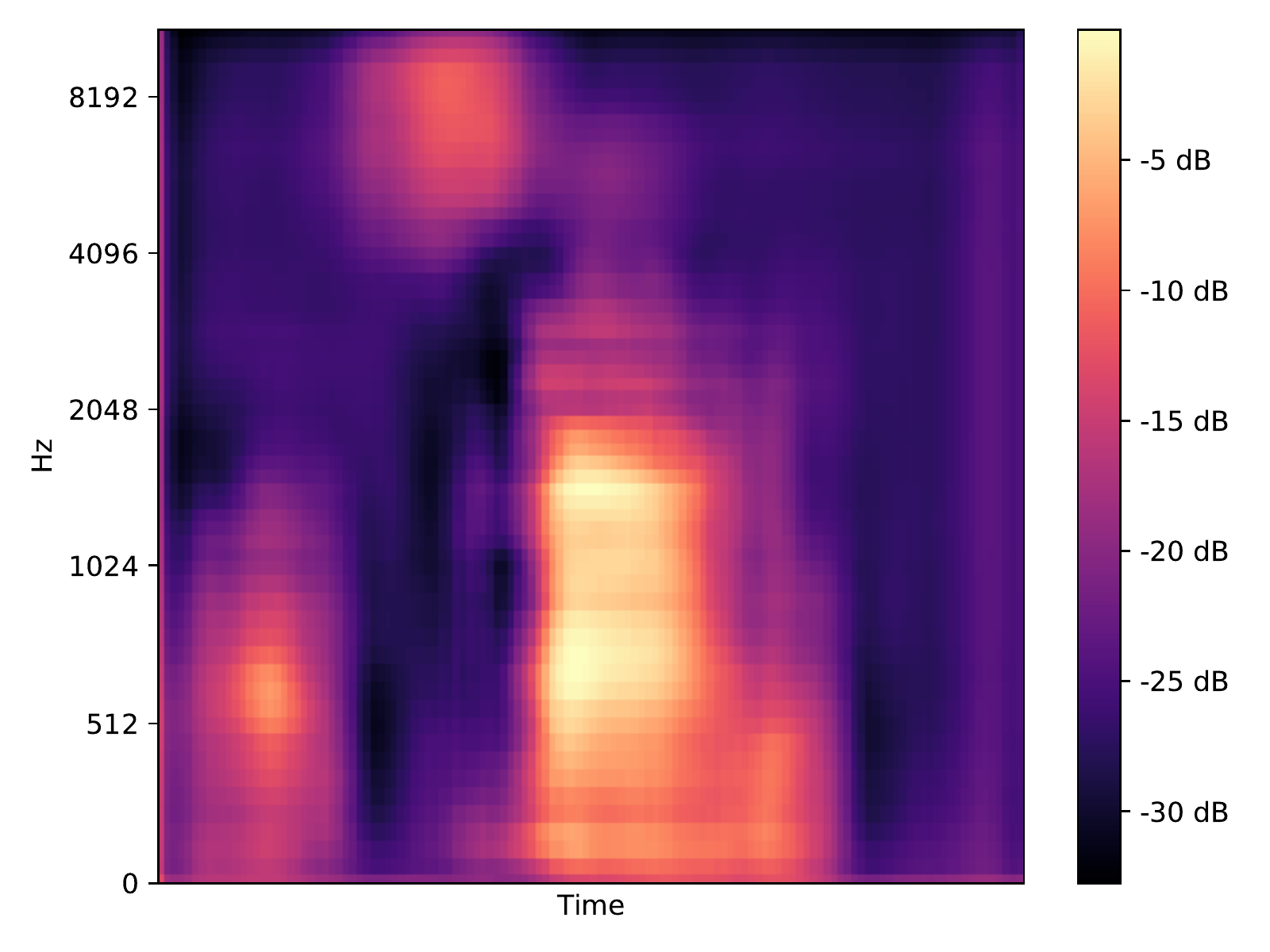}
            \caption[]%
            {{\small PATE-AAE}}
            \label{fig:mean and std of net44}
        \end{subfigure}
        \caption[  ]
        {\small Mel-spectrogram of (a) PATE-GAN; (b) PATE-AAE with a privacy target ($\varepsilon$=0.1), where output command is "right."}
        \label{fig:3}
    \end{figure}

\begin{figure}[ht!]
    \centering
    \includegraphics[width=0.88\linewidth]{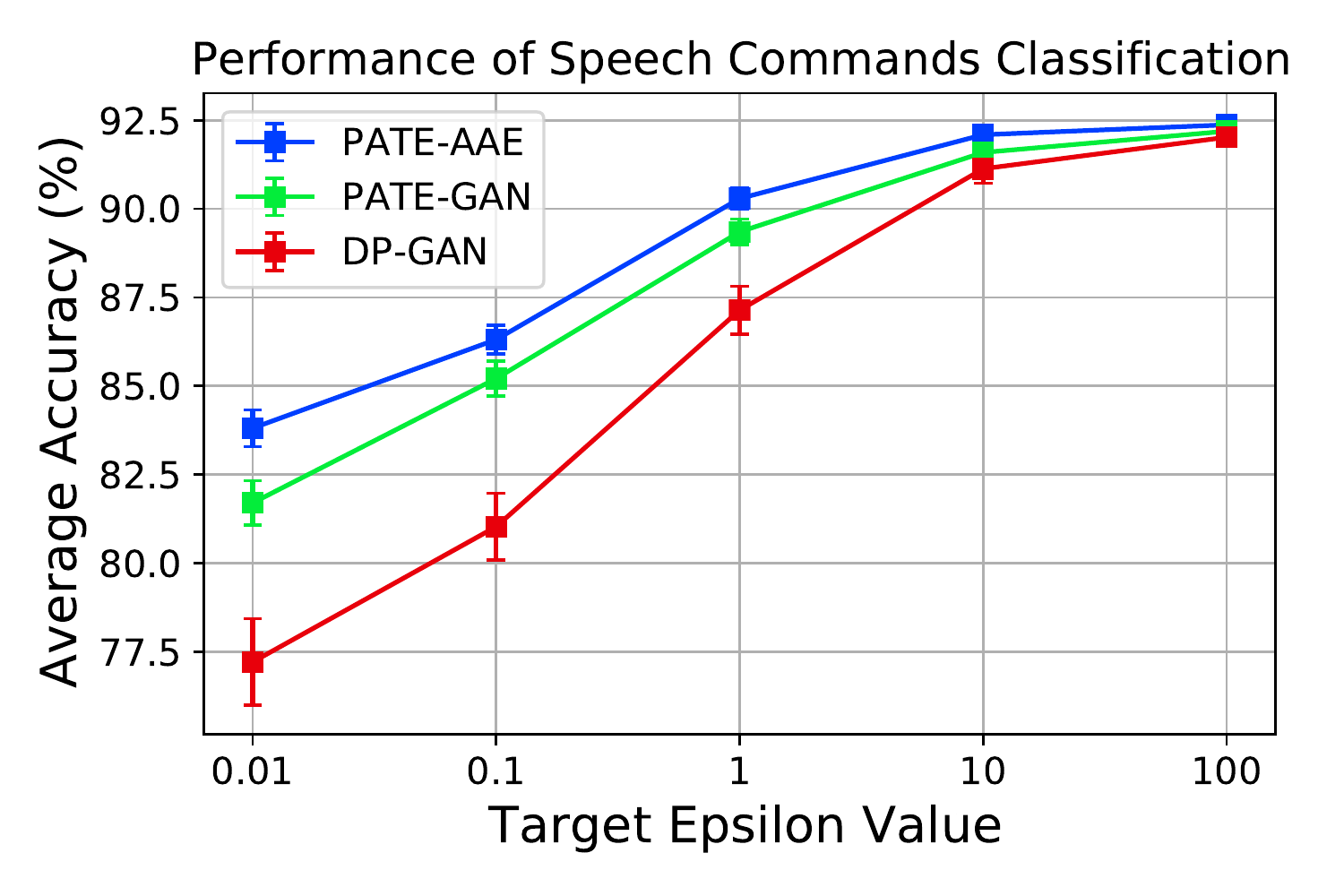}
    \caption{Performance of different privacy-preserving models under four levels of privacy target ($\varepsilon$, with a fixed $\delta$=10$^{-5}$), which refers to the Laplace noise level at scale of $\lambda=\frac{K}{2\varepsilon}$. }
    \label{fig:4}
\end{figure}

\subsection{Results and Performance Analysis}
We have developed two baseline systems, namely  PATE-GAN~\cite{jordon2019pate} and DP-GAN~\cite{xie2018differentially}, whose models are using the same encoder-decoder architecture introduced in Sec. 4.1.(2). First, as shown in Tab.~\ref{tab:1:fid}, PATE-AAE performs best (lowest) FID scores compared with PATE-GAN and DP-GAN, which indicates a good capability of privacy-preserving speech synthesis. According to a visualization presented in Fig.~\ref{fig:3} (a), PATE-AAE demonstrates many detailed acoustic features on its mel-spectrogram, where PATE-GAN's mel-spectrogram at most preserves intensity information. Next, we have investigated the prediction accuracy in terms of different target privacy budgets with the constraint of guaranteeing $(\varepsilon,10^{-5})$-differential privacy. The classification results are averaged over 20 trials to reduce the effect of random parameters initialization.

Fig.~\ref{fig:4} summarises our results. As a fair comparison, a visual inspection of Fig.~\ref{fig:4} reveals that the two baseline systems and the proposed PATE-AAE system attain a similar average classification accuracy with an extremely weak privacy budget ($\varepsilon$=100).
As the constraints on the privacy increases, that is, $\epsilon$ is reduced (the noisy level, $\lambda=\frac{K}{2\varepsilon}$, is increased). Simultaneously, the proposed PATE-AAE approach exemplifies its advantage over both the DP-GAN and PATE-GAN solutions. In particular, PATE-AAE boosts the average accuracy by $2.11\%$ ($\varepsilon$=0.01) and $1.11\%$ ($\varepsilon$=0.1) compared with PATE-GAN; $6.60\%$ ($\varepsilon$=0.01) and $5.32\%$ ($\varepsilon$=0.1) compared with DP-GAN. With a large $\varepsilon$ value, the noise level ($\lambda$) becomes too small; nonetheless, PATE-AAE still attains a slightly better average accuracy (92.37\%) than PATE-GAN (92.19\%) and DP-GAN (92.02\%) baselines. This phenomenon could be due to the aggregation in PATEs ameliorating the negative impact of $\varepsilon$-DP noise and echoing theoretical studies in~\cite{papernot2018scalable, jordon2019pate, chen2020gs}.

\section{Conclusion}
In this paper, we incorporate an adversarial autoencoder into the PATE scheme. We further investigate different privacy-preserving solutions to speech command classification by combining the WaveNet based encoder-decoder structures and classification models together. The proposed PATE-AAE approach shows the best performances in terms of synthetic speech quality scores and classification accuracy. Our future work includes exploring different $\varepsilon$-DP distributed training strategies, such as average gradient aggregation~\cite{chen2020gs} and adversarial training~\cite{yang2020characterizing}, for large vocabulary continuous speech recognition.

\clearpage
\bibliographystyle{IEEEtran}

\bibliography{mybib}

\end{document}